\documentclass[letter]{aa}
\usepackage{newtxtext,newtxmath}
\usepackage[colorlinks,citecolor=blue]{hyperref}

\usepackage[T1]{fontenc}
\usepackage{ae,aecompl}

\usepackage{graphicx}   
\usepackage{amsmath}    

\usepackage{bm}         

\begin{document}

\title{Phase-space distortion as a key to unraveling galactic bar buckling}

\author{
  Viktor D. Zozulia\inst{1,2}
\and 
  Natalia Ya. Sotnikova\inst{1,2}
\and 
  Anton A. Smirnov\inst{2}
}
\institute{
  St. Petersburg State University,
  Universitetskij pr.~28, 198504 St. Petersburg, Stary Peterhof, Russia \\
  \email{n.sotnikova@spbu.ru}
  \and
  Central (Pulkovo) Astronomical Observatory of RAS, Pulkovskoye Chaussee 65/1, 196140 St. Petersburg, Russia
}


\date{Received XXX; accepted YYY}





\abstract
{
For the first time, we investigate the resonant structure of $N$-body galactic bar at the stage of buckling using action-angle variables. We studied the evolution of vertical actions ($J_z$) and angles associated with vertical resonance ($\theta_\mathrm{res}=\theta_z - \theta_R$) for all orbits in the bar. For this purpose, we divided the orbits into types according to the behavior (libration or circulation) of their resonant angle with respect to fixed points $\theta_\mathrm{res}=0$ and $\pi$ (vertical resonance). We show that during buckling, flat bar orbits circulating with increasing $\theta_\mathrm{res}$ transformed into banana-shaped librating orbits (resonant capture) or circulating orbits with decreasing $\theta_\mathrm{res}$ (resonant heating). The orbital transformation is accompanied by an increase in $J_z$ and the formation of a boxy- or peanut-shaped bulge. During buckling, the phase space $J_z - \theta_\mathrm{res}$ undergoes a distortion creating an asymmetry in the position of the fixed points $\theta_\mathrm{res}=0$ and $\pi$ and in banana-shaped orbits near these points. The fixed point $\theta_\mathrm{res}=0$ may disappear completely. This also breaks the symmetry between the orbits, which are captured into resonance or go into circulation with decreasing $\theta_\mathrm{res}$ near $\theta_\mathrm{res}=0$ and $\pi$. At the same time, near $\theta_\mathrm{res}=0$, banana-shaped orbits with low vertical action $J_z$ appear. This reopens the path of orbital transformation through the fixed point, $\theta_\mathrm{res}=0$. The phase space transformation and orbit transformation occur in a coordinated manner and lead to the smoothing of phase space perturbations and restoration of symmetry between orbits.
}

\maketitle



\section{Introduction}
\label{sec:intro}

Recently, there has been a flash of interest in unraveling the nature of buckling, the sudden loss of vertical symmetry by a galactic bar \citep{Collier2020,Sellwood_Gerhard2020,Li_etal2023,Lokas2019,Lokas2025}. This violent buckling episode is observed in many numerical models of disk galaxies (e.g., \citealp{Sotnikova_Rodionov2003,MartinezValpuesta_etal2006,Smirnov_Sotnikova2019,Lokas2019,Collier2020,Kataria2024}) and often precedes a rapid thickening of the bar and the formation of the so-called boxy- or peanut-shaped (BP) bulge  \citep{Combes_Sanders1981}.
\par
In numerical simulations of barred disk galaxies, buckling was first noticed by \citet{Friedli_Pfenniger1990} and then described as a rapid asymmetric protrusion of the bar in numerical simulations of \citet{Raha_etal1991}. \citet{Raha_etal1991} linked it to the fire-hose instability of a thin anisotropic layer. At the same time, \citet{Combes_etal1990,Pfenniger_Friedli1991,Quillen2002} explained the general thickening of the bar by the lifting of orbits captured by or passing through the emerging vertical resonance. \citet{Quillen_etal2014} constructed first- and second-order Hamiltonian models of vertical resonance and associated bar asymmetry during its \text{red}{rapid} thickening with a rapid distortion of the phase space $\theta_\mathrm{res}-J_z$, in which fixed points (corresponding to 3D periodic orbits) for the same value of the Hamiltonian $H$ have different values of $J_z$ for $\theta_\mathrm{res} = 0$ and $\pi$ (a first order resonance). 
\par
Despite the existing investigations concerning the transformation of individual orbits during buckling (e.g., \citealp{MartinezValpuesta_etal2006,Lokas2019,Valencia-Enriquez_etal2023}), there is still no consensus on the nature of buckling in full-fledged $N$-body simulations. \citet{Sellwood_Gerhard2020} insist on fire-hose instability.
\citet{Li_etal2023,Lokas2025} argue for the resonant nature of vertical heating. They focus on the fact that, shortly before buckling, there is a distortion of the Laplace plane in the same direction along the $z$-axis at both ends of the bar. As a result, a periodic force with a frequency of $2(\Omega-\Omega_\mathrm{p})$ (where $\Omega$ is the angular velocity and $\Omega_\mathrm{p}$ is the bar pattern speed) begins to act on the almost flat orbits that oscillate with a high vertical frequency. As a consequence, the orbit undergoes forced oscillations with a vertical frequency of $\omega_z \approx 2(\Omega-\Omega_\mathrm{p})$. In other words, it enters a vertical resonance. Since the perturbing force pushes the orbits at the ends of the bar predominantly in one direction, a vertical asymmetry arises.
\par
In this letter we present a different picture of what occurs during buckling. We associate the symmetric exit of the bar orbits from the midplane with perturbations of $\sim \cos(2 \theta_\mathrm{res})$ (the second-order resonance, \citealp{Quillen_etal2014}), which arise as a consequence of bar formation and its quiet evolution, and create two fixed points (at $\theta_\mathrm{res} = 0$ and $\pi$) in the phase space. However, the culprit of the buckling is asymmetric perturbations, $\sim \cos\theta_\mathrm{res}$, that distort the phase space. If they are strong enough, they allow only one fixed point in the phase space\footnote{$\theta_\mathrm{res}=\pi$ in our model.} (the first-order resonance, \citealp{Quillen_etal2014}). Using the framework of angle-action analysis developed for the self-consistent $N$-body models in our previous papers \citet{Zozulia_etal2024a,Zozulia_etal2024b},  we show for the first time how individual orbits interact with asymmetric phase-space perturbations in the system, and how these interactions result in changes to the overall demographics of orbits. Thus, we present here the first detailed description of buckling in terms of individual orbit types differentiated based on the behavior of their resonant angle.

\section{Numerical model}
\label{sec:nbody}
We used the same numerical model as in~\cite{Zozulia_etal2024a,Zozulia_etal2024b} and refer to the cited papers for more details. The model initially consisted of two components, a pure stellar disk (exponential) and an isotropic spherical dark halo that has a Navarro-Frenk-White-like density profile~\citep{NFW}. The model parameters were presented in the natural system of units, where the gravitational constant $G=1$, the mass of the disk $M_\mathrm{d}=1$, and the initial exponential scale of the disk $R_\mathrm{d}=1$. One time unit would correspond to 13.8 Myr if the total mass of the disk were $M_\mathrm{d}=5\times 10^{10}M_\mathrm{\sun}$ and the scale length were $R_\mathrm{d}=3.5$ kpc. In dimensionless units the initial thickness of the disk is $z_\mathrm{d}=0.05$, where $z_\mathrm{d}$ enters in the sech$^2$" law, the dark halo mass is $M_\mathrm{h}(R<4R_\mathrm{d})\approx1.5$, and the minimum value of the Toomre parameter is $Q(2R_\mathrm{d})=1.2$.
\par
The components evolve in a self-consistent manner until $t=600$ ($\sim 8$~Gyr). A strong and fast bar appears in the center of the disk at about $t=80$. Almost immediately after its formation, the bar begins to thicken. In the time interval from $t=160$ to $t=190$, the bar buckles rapidly and asymmetrically, reaching maximum asymmetry in the vertical direction at $t=180$. By this time, a clearly visible BP bulge develops, which after $t=200$ gradually becomes symmetric. Subsequently, the flat bar and the BP bulge continue to increase in size, capturing new disk orbits and lifting them out of the disk midplane~\citep{Zozulia_etal2024a,Zozulia_etal2024b}. 

\section{Action-angle variables and orbital types}
\label{sec:actions_and_types}
We associated different types of orbits with different areas in the phase space based on the behavior of the resonant angle:
\footnote{\citet{Quillen_etal2014}, considering the general scheme of the phase space ($(\sqrt{J_z}\,\cos{\phi},\,\sqrt{J_z}\,\sin{\phi})$) transformation, employ another designation for the resonant angle $\phi=\theta_z-2(\theta_\varphi-\Omega_\mathrm{p} t)$. However, for the inner Lindblad resonance (ILR, bar orbits) $2\theta_\varphi - \Omega_\mathrm{p} t = 2\theta_\phi \approx \theta_\mathrm{R}$, which means approximate equality $\theta_\mathrm{res} = \theta_z-\theta_\mathrm{R} \approx \phi$.} 
$\theta_\mathrm{res}=\theta_z-\theta_R$.
We calculated the evolution of the non-perturbed, medium-term action-angle variables $(J_\mathrm{R},J_z,\,L_z, \theta_\mathrm{R}, \theta_z, \theta_\phi)$ directly in the $N$-body model, as described in \citep{Zozulia_etal2024a, Zozulia_etal2024b}. 
These medium-term variables change smoothly on the timescale of short-term orbital oscillations, allowing orbital libration and circulation to be traced.
The actions preserve the mean value of instantaneous actions (calculated in an axisymmetric potential using the {\tt{AGAMA}} package, \citealp{agama}) between apocenters or $z$-maxima. Angles and frequencies were calculated consistently such that $\theta_R$ changes by $2\pi$ between apocenters,  $\theta_z$ changes by $2\pi$  between $z$-maxima, and $\theta_\phi$ changes by the azimuthal angle traversed by the particle between the apocenters.
We also used secular values of $J_z$, which are obtained by averaging medium-term values, and thus tracked the orbital evolution on libration timescales. Across this letter we specify which $J_z$ value is used (medium-term or secular).
\par
For each bar particle, we trace the evolution of the resonant angle $\theta_\mathrm{res}=\theta_z-\theta_R$ forward and backward in time, identify four possible modes of its behavior, and thus obtain four types of orbits. These four types of orbits may be described as follows:
\par
    (vCIR$+$) Circulation of the resonant angle with positive angular velocity, $\dot{\theta}_z > \dot{\theta}_\mathrm{R}$ ($\omega_z-\kappa>0$).
\par
    (vCIR$-$) Circulation of the resonant angle with a negative angular velocity, $\dot{\theta}_z < \dot{\theta}_\mathrm{R}$ ($\omega_z-\kappa<0$). This orbit circulates, if $\theta_\mathrm{res}$ passes through at least $2\pi$ in either the positive or negative direction. The notations "up" and "down" of both circulations correspond to the value of the resonant angle in the range $-\pi/2<\theta_\mathrm{res}<\pi/2$ and $\pi/2<\theta_\mathrm{res}<3\pi/2$, respectively. 
\par
    (BAN) Orbits in a vertical inner Lindblad resonance (vILR), whose $\theta_\mathrm{res}$ undergoes at least one libration (revolution) around a stable fixed point near $0$ or $\pi$ (BAN up and BAN down in our notation). Such orbits are typically banana-shaped. 
\par
    (vPAS) Orbits in the process of changing the direction of circulation pass close to the vILR, but do not get stuck in it. The passage, in turn, can occur near $0$ (vPAS up) or near $\pi$ (vPAS down). 
\par
A detailed description of the calculation of the action-angle variables and the determination of different orbital types (including ILR orbits) based on the changes in resonant angle is given in Appendix~A of \citet{Zozulia_etal2024b} and in Appendix~\ref{sec:appendix_types} of this letter.

\section{Demographics of orbits}
\label{sec:distortion}
\begin{figure*}
\centering
\includegraphics[width=1.0\linewidth]{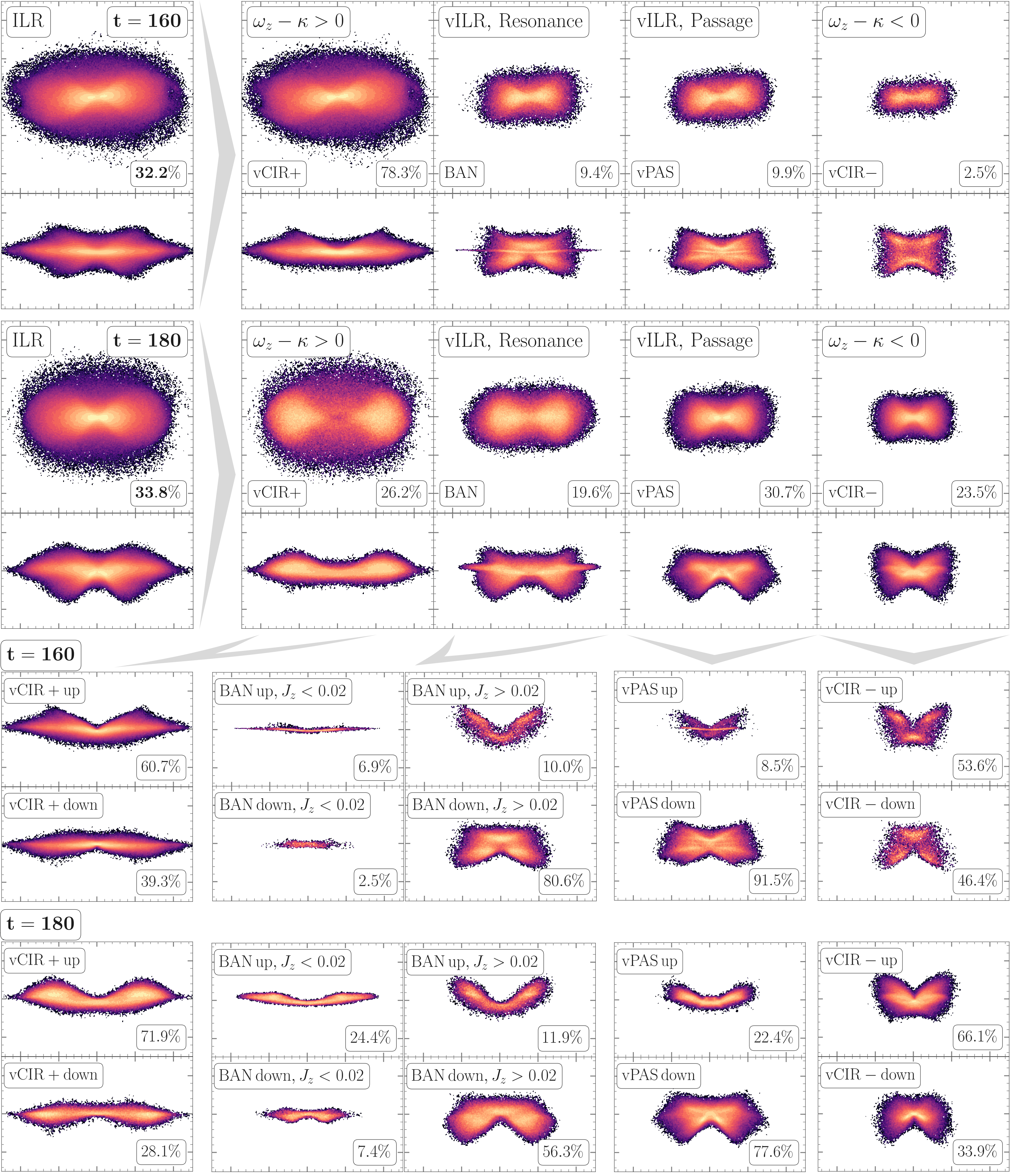}\\
\caption{Dynamic decomposition of the bar (ILR) at the beginning of buckling ($t=160$, \textit{two top row}) and at the peak of buckling ($t=180$, \textit{two bottom row}). For each time moment, the five leftmost snapshots on the $xy$-plane, covering $[-2.5, 2.5] \times [-2.5,2.5]$, and on the $xz$-plane, covering $[-2.5, 2.5] \times [-1.5,1.5]$, are shown. \textit{From left to right}: All bar orbits (ILR); orbits with increasing resonant angle, $\theta_\mathrm{res}=\theta_z - \theta_\mathrm{R}$ (vCIR$+$); orbits in the vILR (BAN); orbits passing through it (vPAS); and orbits with decreasing $\theta_\mathrm{res}$ (vCIR$-$). The next five snapshots show the same orbits on the $xz$-plane, but separated into "up" (near $\theta_\mathrm{res}=0$) and "down" (near $\theta_\mathrm{res}=\pi$) types, and into secular $J_z < 0.02$ (very flat) and $J_z>0.02$ (thick) for BAN orbits. For more information on the classification of these types, see Sec.~\ref{sec:actions_and_types} and Appendix~\ref{sec:appendix_types}.}
\label{fig:snapshots}
\end{figure*}

Figure~\ref{fig:snapshots} shows how the orbital types in the bar change in relation to $\theta_\mathrm{res}$ over the time interval from $t=160$ to $t=180$ (moment of the highest vertical asymmetry of the bar). The amplitude of the bar during buckling in our model practically does not change (see Fig.~3, left plot in \citealp{Smirnov_Sotnikova2018}, green line). The fraction of particles in the bar of all disk particles from $t=150$ to $t=200$ changes slightly, from 32.3\% to 33.7\%. It then slowly increases to 42.1\% by $t=300$. The vertical structure of the bar at $t=160$ is almost symmetric (Fig.~\ref{fig:snapshots}, ILR). Most of the bar's orbits are nearly flat (vCIR$+$, 78.3\%), but among these orbits there is a bias toward $\theta_\mathrm{res} = 0$. This bias is compensated for by the reverse bias of the BAN orbits and the vPAS orbits toward $\theta_\mathrm{res} = \pi$ ("down" types). 
\par
The picture changes dramatically at $t=180$. By this time, almost two thirds of the vCIR$+$ orbits are transformed into other types, and although the vCIR$+$ orbits are still dominated by orbits near $\theta_\mathrm{res} = 0$ (18.8\% of the total orbits in the bar versus 7.4\% of the vCIR$+$ orbits near $\theta_\mathrm{res} = \pi$), a strong vertical asymmetry is visible in the combined snapshot. The asymmetry is due to the large percentage of the BAN down and vPAS down orbits near $\pi$ (36.3\% of the total number of the orbits in the bar). These orbits have significantly greater $J_z$ than the vCIR$+$ orbits. An additional vertical skew is associated with the appearance of very flat (secular $J_z < 0.02$) BAN up orbits ($\theta_\mathrm{res} = 0$) that are greatly extended along $R$.
Although the total number of all types of orbits near $\theta_\mathrm{res} = 0$ (48.3\%) is almost the same as that near $\theta_\mathrm{res} = \pi$ (51. 7\%), orbits near $\theta_\mathrm{res} = \pi$ are dominated by those that are in the process of transformation and have high values of $J_z$. This determines the vertical asymmetry.
\par
\begin{figure*}
\centering
\includegraphics[width=1.0\linewidth]{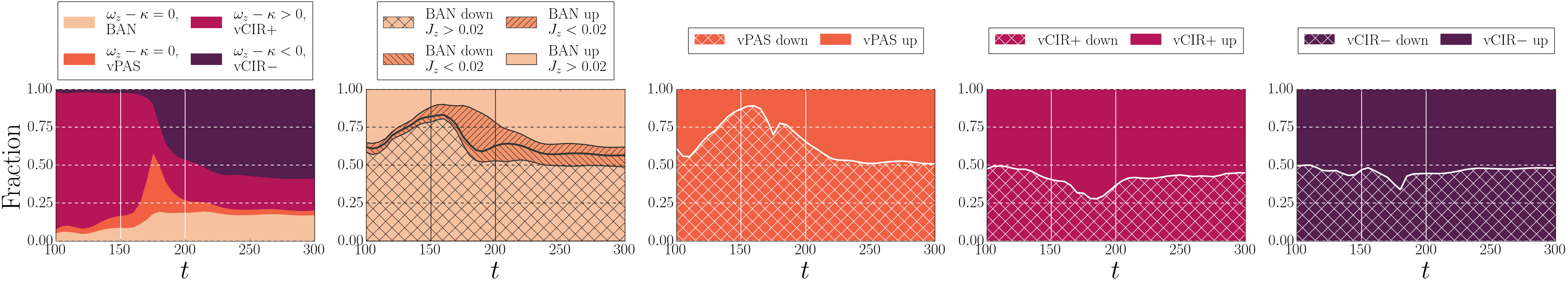}
\caption{The evolution of the fraction between different types of orbits from $t=100$ to $t=300$. In all plots, the buckling stage $(t=150-200)$ is highlighted by vertical lines. \textit{From left to right:} among orbits in ILR (the bar), among ILR orbits captured in vILR (BAN), among ILR orbits passing through vILR (heating), among ILR orbits with increasing resonant angle $\theta_z-\theta_\mathrm{R}$ (vCIR$+$, flat orbits) and among ILR orbits with decreasing resonant angle $\theta_z-\theta_\mathrm{R}$ (vCIR$-$). The different types of orbits in plots, starting with the second, are shown in the same colors used in the first plot. All types correspond to those shown in Fig.~\ref{fig:snapshots}.}
\label{fig:hist_t_in_resonance}
\end{figure*}

Figure~\ref{fig:hist_t_in_resonance} shows all of the noted features in changes in the percentages of orbits of different types. The sharp bias of the PAS down orbits with respect to the PAS up ones shortly before buckling and the sharp increase in all the PAS orbits at the moment of buckling are especially visible. Additionally, we note that by $t=300$, the symmetry between the orbits near $\theta_\mathrm{res} = 0$ and $\pi$ is practically restored for all types of orbits, although a slight asymmetry still remains.

\section{Phase-space distortion and portals for transferring orbits}
\label{sec:evolution}
In \citet{Zozulia_etal2024b} we presented typical phase trajectories of individual orbits at late stages of bar evolution, after $t=400$.
Here, to explain the bias in the number of orbits of different types near $\theta_\mathrm{res} = 0$ and $\pi$, we construct phase portraits of orbital ensembles during buckling, which give an idea of the structural evolution and distortions of the phase space.\par
We developed a method for obtaining sketches of phase portraits on the planes $(\theta_{\mathrm{res}}=\theta_z-\theta_\mathrm{R},\,J_z)$ and $(\sqrt{J_z}\,\cos{\theta_\mathrm{res}},\,\sqrt{J_z}\,\sin{\theta_\mathrm{res}})$ and tracked their evolution directly in the $N$-body model. To do this, we found orbits that align along the major axis of the bar and have the same fixed value of the Jacobi integral, $H_J$, and track the evolution of their $\theta_{\mathrm{res}}$ and $J_z$.
This approach enabled the reduction of the 6D phase space of the bar to a 2D one, thus facilitating a more detailed study. Here, we used medium-term $J_z$ and $\theta_{\mathrm{res}}$ so as not to miss the libration of the orbits.
More detailed information on the construction of phase portraits can be found in Appendix~\ref{sec:appendix_PP}.
\par
Figure~\ref{fig:phase_port} shows the evolution of the phase portraits described above for different values of $H_J$ (higher modulus values correspond to more central orbits) during and after buckling. The layered structure of the phase space can be observed on the plane $(\theta_\mathrm{res}, J_z)$ at each moment of time and for all $H_J$. We can see that at $t=160$ vCIR$+$ (the bottom orbit layer) and BAN down (the upper layer around $\theta_{\mathrm{res}}=\pi$) orbits occupy most of the phase space. The area size of the BAN up orbits is negligibly small. For given values of $H_J$, there is not even a stable fixed point near $\theta_\mathrm{res}=0$. The areas occupied by vCIR$-$ orbits (the uppermost layer), flat BAN up orbits (the new lowest layer), and non-flat BAN up (the central layer around $\theta_{\mathrm{res}}=0$) increase significantly by $t=180$ and $200$. All of this fits perfectly with the picture that we saw after performing the dynamic decomposition of the bar (Fig.~\ref{fig:snapshots}). Also, on maps $(\sqrt{J_z}\,\cos{\theta_\mathrm{res}},\,\sqrt{J_z}\,\sin{\theta_\mathrm{res}})$ we can see that flat BAN up and vCIR$+$ orbits have a common origin and circulate around the stable point that is offset from the origin. It should be added that in the absence of disturbances associated with the angle $\theta_\mathrm{res}$, these maps show only contours concentrated around the origin. When there are strong perturbations of $\sim \cos\theta_\mathrm{res}$ ($t=160-200$), the maps show additional contours around a point offset from the origin.
\par
In a self-consistent way, the transformation of orbits in a distorted phase space transforms the phase space itself, which in turn changes the transformation paths of individual orbits. Figure~\ref{fig:portals} shows possible paths of orbit evolution under such a transformation. Orbits move from vCIR$+$ to BAN up or down or to vCIR$-$ through vPAS up or down; from non-flat BAN up or down to vCIR$-$; and from flat BAN up to vCIR$+$, non-flat BAN up, or vCIR$-$, depending on a phase space structure. For example, at $t=160\,,180$ for $H_J=-2.2$, as seen in Fig.~\ref{fig:phase_port} and Fig.~\ref{fig:portals}, there is no libration area near the stable point $(\theta_z-\theta_R=0,J_z\approx0.1)$, but there is a large area near $(\theta_z-\theta_R=\pi,J_z\approx0.1)$. Thus, the orbits move from vCIR$+$ to BAN down or vCIR$-$ through vPAS down, or to flat BAN up. Later, when a large number of vCIR$+$ orbits have transformed, conditions are created for a second stable point at $\theta_\mathrm{res}=0$ (later $t=200$). From this moment, the transformation of orbits begins to occur through both portals ($\theta_\mathrm{res}=0$ and $\pi$), and the phase space gradually symmetrizes (Fig.~\ref{fig:phase_port}, $t=300\,,400$). 

\begin{figure*}
\centering
\includegraphics[width=1.0\linewidth]{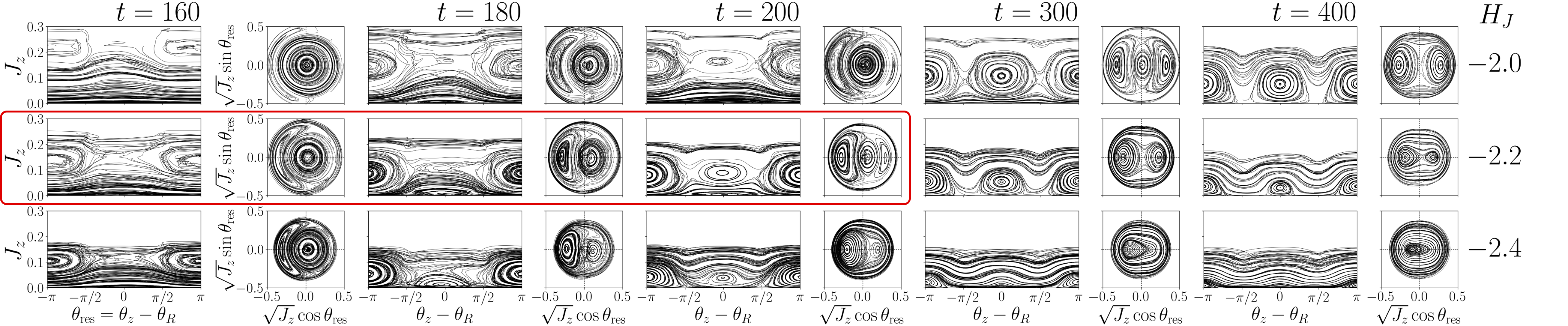}
\caption{Evolution of instantaneous phase portraits on the $(\theta_z-\theta_\mathrm{R}, J_z)$ plane $[-\pi,\pi] \times [0,0.3]$ and on the $\sqrt{J_z}\left(\cos{(\theta_z-\theta_\mathrm{R}) },\sin{(\theta_z-\theta_\mathrm{R}})\right)$ plane $[-0.5,0.5] \times [-0.5,0.5]$ for three different $H_J$ values. See Appendix~\ref{sec:appendix_PP} for a detailed description.
The red box contains the phase portraits, which are discussed separately.}

\label{fig:phase_port}
\end{figure*}

\begin{figure*}
\centering
\includegraphics[width=1.0\linewidth]{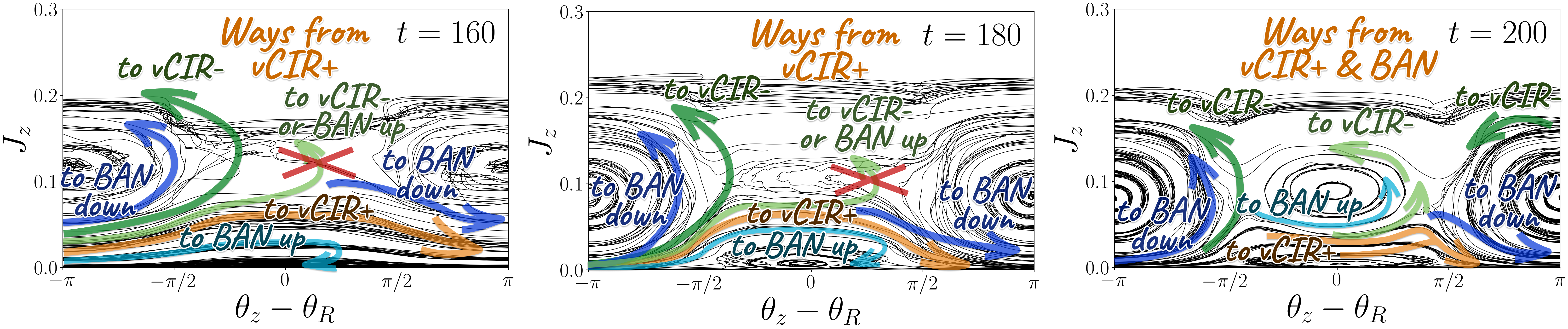}
\caption{Phase portraits on the $(\theta_z-\theta_\mathrm{R}, J_z)$ plane for orbits along the major axis of the bar with fixed $H_J=-2.2$. These are the same portraits contained in the red box in Fig.~\ref{fig:phase_port}. Portraits are numerically obtained at three points in time $(t=160,\,180,\,200)$. The bold arrows show the possible and forbidden (for a given $H_J$) paths of transformation of vCIR$+$ orbits (two left panels) and vCIR$+$ with BAN orbits (right panel) during the evolution of the phase space.}
\label{fig:portals}
\end{figure*}

\section{Discussion}
\label{sec:discussion}
The term buckling is employed to denote not only the rapid protrusion of the bar from the midplane, but also the asymmetric protrusion in the vertical direction\citep{Raha_etal1991,MartinezValpuesta_etal2006,Sellwood_Gerhard2020}. This is believed to result in the formation of BP bulges. However, under certain conditions, BP bulges can also form completely symmetrically \citep{Smirnov_Sotnikova2018,Smirnov_Sotnikova2019,Sellwood_Gerhard2020,McClure_etal2025}. This difference in the morphology of the process forces researchers to look for different mechanisms that can facilitate the exit of the bar orbits from the midplane in both symmetric (resonant heating or resonant trapping) and asymmetric (instability) cases \citep{Sellwood_Gerhard2020}.
\par
In this letter we want to emphasize that after the bar emergence and during its further evolution, the shape of the phase space changes due to perturbations of the Hamiltonian of the system.
We leave aside the reason for these perturbations, but when the Hamiltonian of the system changes, both resonance trapping and heating can occur \citep{Quillen2006}. This in turn leads to a redistribution of orbits in phase space. The system responds by further changing the Hamiltonian. Transformation of orbits and phase space occurs simultaneously and in a self-consistent manner.
\par
Our simulations show that during buckling the asymmetric exit of orbits from the midplane is caused by perturbations of the Hamiltonian, both first- and second-order \citep{Quillen_etal2014}, with both heating and trapping occurring simultaneously. If the amplitude of first-order Hamiltonian perturbations is large, the system admits only one fixed point in a wide range of values of the Jacobi integral (Fig.~\ref{fig:phase_port}), and the exit of orbits from the midplane are asymmetric. Moreover, judging by the spike in the fraction of vPAS orbits in the first plot of Fig.~\ref{fig:hist_t_in_resonance}, there is rather resonant heating during buckling. When first-order perturbations are absent or small (as in the case of a central concentration of mass, \citealp{Smirnov_Sotnikova2019,McClure_etal2025}), the transformation of the orbits occur through two fixed points -- that is, almost symmetrically --  and the resonant capture dominates. \citet{Zozulia_etal2024b} show that at this stage the orbits can remain in vertical resonance for a long time.
\par
Although \citet{Li_etal2023,Lokas2025} consider buckling to be a resonant phenomenon, we cannot agree with their interpretation of the orbital transformation. In their picture, the initial bisymmetric distortion of the Laplace plane at the ends of the bar (upward bending) leads to forced oscillations of the orbits with a vertical frequency of $\nu\approx2(\Omega-\Omega_\mathrm{p})$. As a result, an excess of BAN up orbits is formed. Later, the distortion winds up and causes the transformation of banana-like orbits into pretzel-like one and gives rise to BAN down orbits.
\par
In our picture, with the same initial distortion of the Laplace plane (due to the excess of vCIR$+$ up orbits), the role of a portal for transferring orbits to other regions of the phase space is played by a fixed point $\pi$, around which the BAN down orbits are concentrated from the very beginning. We consider that the transformation of orbits near this point occurs in the same way as in the second-order Hamiltonian model, with two fixed points \citep{Quillen_etal2014}. Figure~\ref{fig:phase_port} shows that the height ($J_z$) of the fixed point $\pi$ for the same value of the Jacobi integral decreases over time as the bar slows down, the Hamiltonian of the system changes, and the orbits are redistributed. The dropping down of fixed points reflects the phenomenon of separatrix shrinking \citep{Quillen_etal2014}. In this case, the librating BAN orbits, which were initially captured from the vCIR$+$ region, can cross the separatrix, transforming into vCIR$-$ orbits with high $J_z$ values. We proceed from considerations of preserving the phase volume (Liouville's theorem). As a result, if the orbits leave the circulation mode (vCIR$+$), the phase-space volume of the vCIR$+$ area decreases, and the volume of the areas where the orbits transfer increases. After a significant reduction in the area occupied by vCIR$+$ orbits, the conditions for the appearance of a second fixed point near zero arise.
\par
We would also like to note that our classification of orbits does not contradict other classifications based, for example, on the ratio of vertical to radial oscillation frequencies, $\omega_z/\kappa$. The value of $\omega_z/\kappa$ is often used to distinguish orbits from which different parts of the BP bulge are assembled (e.g., \citealp{Portail_etal2015b,Parul_etal2020}). Our classification, based on the resonance angle behavior, paints a picture of the phase space in broad strokes. In it, for example, all orbits with a low frequency ratio fall into the region behind the separatrix, i.e., the region of vCIR$-$ orbits, although among these there may be beautiful periodic orbits, such as "brezels" \citep{Portail_etal2015b}.
\section{Conclusions}
\label{sec:conclusion}
\citet{Zozulia_etal2024b} studied the long-term and gradual evolution of the entire ensemble of orbits in the bar.
Here, we focus on the early stages of a bar orbit evolution in the vertical direction, when all changes occur rapidly (buckling). Our analysis shows that differences in the behavior of the resonant angle correspond to different orbits in a phase space, which, in turn, create different structures in the $xyz$-space. During buckling, the perturbed phase space (the Hamiltonian of the system) evolves, and orbits change their types. The transformation of orbits leads to a transformation of phase space, which in turn leads to a redistribution of orbits with different behaviors in the resonant angle, $\theta_\mathrm{res}$. This happens in a self-consistent manner. As a result, the vertical shape of the bar changes.
\par
Our main conclusions about what occurs during buckling are as follows:
\par
    1. The phase space of the bar is arranged as a layer cake. The lower layer, with small $J_z$ (flat orbits), consists of circulating orbits with an increasing resonant angle (vCIR$+$). Next is the intermediate layer (the "filling"), containing banana-shaped orbits (BAN up and BAN down) that librate around $0$ or $\pi$ . Finally, the top layer contains orbits with large $J_z$ that reduce the resonant angle (vCIR$-$). During the vertical growth of the bar, the shape of these layers changes, and the orbits are redistributed between them.
    An orbit can move from the lowest layer (vCIR$+$) to the middle layer (BAN) via resonant capturing, stay there for a while, and then move to the uppermost layer (vCIR$-$), or it can pass directly to vCIR$-$ (vPAS) via resonant heating. In the case of a quiet evolution, the phase space ($\theta_\mathrm{res}, J_z)$
    of the bar is symmetric and changes slowly. Orbits move smoothly from one layer to another \citep{Zozulia_etal2024b} as the fixed points in phase space are lowered. The libration orbits either fall with them or gradually move to the upper layer behind the separatrix.
\par
    2. During buckling, the phase space ($\theta_\mathrm{res}, J_z)$ is asymmetric, which means that the transfer of orbits to the higher layers is also asymmetric (Fig.~\ref{fig:phase_port} and~\ref{fig:portals}). 
\par
    3. At the beginning of buckling, there are practically no librating orbits near $\theta_\mathrm{res}=0$
    (BAN down), and the orbits move from vCIR$+$ to BAN up or vCIR$-$ around $\theta_\mathrm{res}=\pi$ (vPAS down). A new lowest layer of flat, librating orbits appears near $\theta_z-\theta_\mathrm{R}=0$. This layer is associated with the curvature of the minimum potential surface (Laplace plane) and exists as long as this curvature exists. Subsequently, a libration region forms at $\theta_\mathrm{res}=0$ and high $J_z$.
    Orbits begin to be trapped in it, and a new transfer portal for vCIR$+$ orbits opens around $\theta_\mathrm{res}=0$ (vPAS up). Gradually, during this transfer, the vCIR$+$ layer becomes thinner, the flat librating orbits disappear, and the phase space becomes symmetric. At this stage, the replenishment of the vCIR$+$ layer occurs at new orbits join the bar while its amplitude grows.

\begin{acknowledgements} 
We acknowledge financial support from the Russian Science Foundation, grant no. 24-22-00376. 
We are also deeply grateful to an anonymous reviewer whose comments not only contributed to a much improved presentation of our results, but also generated a wide range of ideas for future research.
\par
We acknowledge the use of the~\texttt{AGAMA}~\citep{agama} and ~\texttt{mpsplines}~\citep{Ruiz_Jose2022} \texttt{python} packages, without which the present work would not be possible.
\end{acknowledgements}

\bibliographystyle{mnras}
\bibliography{aa54837-25} 

\begin{appendix}
\section{Orbital types}
\label{sec:appendix_types}
In a pendulum model, three modes of its behavior can be distinguished depending on the angle $\phi$ from the vertical, i.e., libration around a stable point and circulation with a decrease/increase in the angle \citep{Lichtenberg_Lieberman_1992}. These modes localize different regions of the phase space. The perturbed Hamiltonian dynamics near a resonance\footnote{When reducing a Hamiltonian system to a resonant one, averaging over fast angles occurs. Our procedure for calculating the action and angles already includes averaging over radial $\theta_\mathrm{R}$ and vertical $\theta_z$ angles.}
can be reduced to a first approximation to the pendulum model, where $\phi$ is equal to the resonant angle $\theta_{\mathrm{res}}$ \citep{Lichtenberg_Lieberman_1992, Binney_2020}. By identifying the three modes of $\theta_\mathrm{res}$ behavior, one can determine in which region of the phase space the orbit is located, at least for regular or sticky chaotic orbits that can change their behavior. It should be noted that, unlike classical mechanics systems, $N$-body models evolve by changing the potential (sometimes quite rapidly) and, therefore, the phase space. This means that orbits can change their behavior relative to the resonant angle. For this reason, we introduce an intermediate type of orbit behavior --- ``passage'' --- for orbits that pass from one circulation to another without getting stuck in resonance.
\par
Obviously, the above reasoning applies to any resonant angle $\theta_\mathrm{res}$. The next step is to address the technical aspects of the matter. How can we identify the specific orbital types? To begin with, we must at least roughly understand the properties of the system: the position of the stable points and the boundaries within which libration regions may be located. In this Letter we consider two resonant angles $2\theta_\phi - \theta_R$ and $\theta_z - \theta_R$. The stable points of the first of them are $0$ for $x1$ orbits (ILR in our notation) and $\pi$ for $x2$ orbits, the libration region can be within $\pm \pi$ relative to these stable points. Similarly, the stable points of the second resonant angle are $0$ or $\pi$. The exact resonances correspond to banana-shaped orbits, the tips of which are directed ``up'' and ``down'', respectively.  
\par
It follows that the orbital type can be determined by the number of times the resonant angle takes a value equal to the angle of a fixed point and by whether it extends beyond the libration region.
\citet{Zozulia_etal2024b} we used a similar approach, but in the current study, we improved the methodology. 
If the system has stable points at $\theta_\mathrm{res} = 0\pm \pi$ and $\pi \pm \pi$, then the identification of types is as follows.
\par
    1, 2. Circulation of the resonant angle $\theta_\mathrm{res}$ with positive $\dot{\theta}_\mathrm{res}>0$ or negative $\dot{\theta}_\mathrm{res}<0$ angular velocity. In this case, the resonant angle $\theta_\mathrm{res}$ successively takes values that are multiples of $\pi$ times in ascending or descending order, respectively.
    \par
    3. Libration around a stable point. $\theta_\mathrm{res}$ takes the value $0$ or $\pi$ at least three times without going beyond the boundaries of $\pm \pi$. Thus, the orbit makes at least one revolution around a stable point in phase space. The orbit enters the resonance when it has $\Delta \theta_{12}$ left to pass to the stable point, where $\Delta \theta_{12}$ is the maximum deviation of the resonant angle relative to the stable point between the first and second passages through it. In fact, this represents the amplitude of the libration angle. The moment of exit from resonance is determined similarly. 
    \par
    4. Passage through the resonance is determined in the same way as libration, but $\theta_\mathrm{res}$ takes the value $0$ or $\pi$ only twice. Formally, the orbit enters resonance, but does not make a full revolution around a stable point in phase space and changes the direction of its circulation. The initial and final times of passage are defined similarly to the libration.
\par
Applying these criteria to the resonant angle $2\theta_\phi - \theta_\mathrm{R}$, we distinguish orbits in inner Lindblad resonance (ILR, x1), orbits with negative angular velocity $\Omega - \kappa/2 < \Omega_p$, and all other orbits (x2, Residuals). We also divide the orbits in circulation into central and disk ones by the minimum distribution of the $z$-component of angular momentum $L_z$ ($L_z^0$). 
\par
In this paper, ILR orbits are divided by the behavior of the resonant angle $\theta_z - \theta_\mathrm{R}$ and are briefly described in Sect.~\ref{sec:actions_and_types}. The first and second types correspond to circulations with positive and negative angular velocity (vCIR$+$ and vCIR$-$), the third and fourth types correspond to librations (BAN) and passage (vPAS) near a stable point.

\section{Sketches of phase portraits}
\label{sec:appendix_PP}
The shape of phase space and the distribution of orbits within it completely determine the structure of a galaxy. However, the main challenge in studying galaxies in this way arises from the 6-dimensionality of this space. 
For a qualitative understanding of the features of the phase space, we need to reduce the dimensionality of the problem. Using the action-angle variables $(J_\mathrm{R}, J_z, L_z, \theta_\mathrm{R}, \theta_\mathrm{\phi}, \theta_z)$ helps us with this. 
\par
In \cite{Zozulia_etal2024a,Zozulia_etal2024b} we describe the method for calculating the averaged action-angle variables that at least eliminates short-period variations of actions, frequencies, and angles. 
Thus, although not quite precisely (which, as follows from classical mechanics, requires a full Fourier transform), we eliminate the fast variables (angles $\theta_R$ and $\theta_z$). This means that away from resonances we pass to three integrals of motion and a three-dimensional phase space. The problem becomes more complicated and interesting near the resonances.
\par
We showed in \cite{Zozulia_etal2024a} that using averaged action-angle variables (medium-term) allows us to reduce the number of dimensions to 4 near ILR and vILR. In this case, we assume that $\dot{\theta}_\mathrm{R}, \dot{\theta_z} \gg \dot{\theta_z}-\dot{\theta}_\mathrm{R}, 2\dot{\theta}_\phi-\dot{\theta}_\mathrm{R}$, and then with a fixed value of the adiabatic invariant $J_v = J_\mathrm{R} +L_z/2 + J_z$ or Jacoby integral $H_J$ it is enough to consider the 4-dimensional space $(L_z, \theta_1 = 2\theta_\phi - \theta_\mathrm{R}, J_z, \theta_2 = \theta_z-\theta_\mathrm{R})$. In turn, if we consider orbits exactly along the bar, then we reduce the number of dimensions to two, implicitly reducing one of the equations of motion to the form $\dot{\theta_1}(L_z, J_z, \theta_1=0, \theta_2) = 0$. This means that, although qualitatively, it will allow us to explore the shape of phase space $(\theta_z-\theta_R, J_z)$ and its evolution.
\par
To investigate the phase space $(\theta_z-\theta_R, J_z)$ we sketch the phase portraits. The algorithm for constructing them is divided into 2 steps. The first step is to search for orbits extended along the major axis of the bar. To do this, we solve a one-dimensional optimization problem. We launch orbits in a frozen rotating potential with coordinates $(x, y=0, z=z_0, v_x=0, v_y=H_J(x,z_0), v_z=0)$, which correspond to $(\theta_\mathrm{R}=0, \theta_z=0, \theta_\phi=0)$. Here, $z = z_0$ and the Jacobi integral $H_J$ are fixed. We need to find the value of $x$ such that the apocenters of the orbits lie on the major axis of the bar. To do this, we trace the evolution of the orbit for a given time ($200$ units, usually), and minimize the sum of the squares of the $y$-coordinate at the apocenter $\sum{y^2(t_\mathrm{apo})}\rightarrow \min$. The second step is to calculate the averaged medium-term $\theta_z$, $\theta_\mathrm{R}$ and $J_z$ of the resulting, and plot the phase portrait $(\theta_z-\theta_\mathrm{R}, J_z)$ of this orbit with different initial values of $z_0$.
\par
It should be noted that only a part of the orbits in the bar is aligned exactly along its major axis, while most of them librate around it. The behavior of these orbits will be complicated as their phase space is four-dimensional. 
However, we believe that the general pattern of orbital evolution should be preserved. We expect only to see the splitting of the separatrix and the formation of a chaotic layer near it, which can accelerate the transition of orbits between different types.
\end{appendix}

\label{lastpage}
\end{document}